\documentclass[preprint,aps,amsmath,amssymb,pacs]{revtex4}
\usepackage{graphicx}
\usepackage{epsfig}
\usepackage{amssymb}
\usepackage{graphicx}
\usepackage{dcolumn}
\usepackage{bm}
\newcommand{\ba}{\begin{array}}
\newcommand{\ea}{\end{array}}
\def\br{\begin{eqnarray}}
\def\er{\end{eqnarray}}
\def\be{\begin{equation}}
\def\ee{\end{equation}}

\def\({\left(}
\def\){\right)}

\begin{document}


\title{A 125 GeV Scalar Boson and $SU(N_{TC})\otimes SU(3)_{{}_{L}}\otimes U(1)_{{}_{X}}$ models}

\author{A.~Doff$^1$ and A.~A.~Natale$^{2,3}$}
\affiliation{$^1$Universidade Tecnol\'ogica Federal do Paran\'a - UTFPR - DAFIS, Av. Monteiro Lobato Km 04, 84016-210, Ponta Grossa - PR, Brazil \\
$^2$Centro de Ci\^encias Naturais e Humanas, Universidade Federal do ABC, 09210-170, Santo Andr\'e - SP, Brazil \\
$^3$Instituto de F\'{\i}sica Te\'orica, UNESP, Rua Dr. Bento T. Ferraz, 271, Bloco II, 01140-070, S\~ao Paulo - SP, Brazil}

\date{\today}

\begin{abstract}

We verify that $SU(N)_{{}_{TC}}\otimes SU(3)_{{}_{L}}\otimes U(1)_{{}_{X}}$ models, where the gauge symmetry breaking is totally dynamical and promoted by the non-Abelian technicolor (TC) group and the strong Abelian interactions, are quite constrained by the LHC data. The theory contains a $T$ quark self-energy involving the mixing between the neutral gauge bosons, which introduces the coupling between the light and heavy composite scalar bosons of the model. We determine the lightest scalar boson mass for these models from an effective action for composite operators, assuming details about the dynamics of the strong interaction theories. Comparing the value of this mass with the ATLAS and CMS observation of a new boson with a mass $M_{\phi} \sim 125$GeV and
considering the lower bound determined by the LHC Collaborations on the heavy neutral gauge boson $(Z^\prime)$ present in these models , 
we can establish constraints on the possible models. For example, if  $SU(N)_{{}_{TC}}\equiv SU(2)_{{}_{TC}}$, with technifermions in
the fundamental representation, the model barely survives the confrontation with the LHC data.

\end{abstract}


\maketitle

\section{Introduction}

\par The Standard Model (SM) of electroweak interactions is in excellent agreement with the experimental data and has explained
many features of particle physics throughout the years. Despite its success there are some points in the model as, for instance, the
enormous range of masses between the lightest and heaviest fermions and other peculiarities that could be better explained at a deeper level
assuming the introduction of new fields or symmetries.

\par Recently the ATLAS and CMS collaborations reported the observation of a new boson with a mass $M_{\phi} \sim 125$GeV which is suspected
to be the SM Higgs boson. The current data on this boson diphoton event rate exhibit the signal strength about (1.5 - 2) larger than the one expected for the standard model Higgs boson. There are already many SM extensions trying to explain this possible enhancement of the $\gamma \gamma$ decay.
In particular, the increase of this decay rate is natural in the context of a 3-3-1 model \cite{felice1, frampton} and its alternative version with exotic leptons\cite{tonasse}, due to the presence of an extra charged vector boson and a doubly charged one as discussed in Ref.\cite{PA}.

\par  This class of models predicts interesting new physics at TeV scale \cite{trecentes} and addresses some fundamental questions that cannot be explained in the framework of the Standard Model\cite{dp1,dp2}. These models also contain a set of fundamental scalar bosons, with many
parameters and clearly suffering from the problems of naturalness and hierarchy \cite{sus,tu}. However in Refs.\cite{Das,331-din1} it 
was suggested that the gauge symmetry breaking in some versions  of the 3-3-1 model \cite{tonasse} could be promoted dynamically, 
because at the scale of a few TeVs  the $U(1)_X$ coupling constant becomes strong and the exotic quark $T$  that appears in the model forms a condensate  breaking  $SU(3)_{{}_{L}}\otimes U(1)_X$ to the SM electroweak symmetry. This is a very interesting feature and peculiar to this class
of models. Unfortunately the SM gauge symmetry still remains intact, and the nice characteristics of the model could be missed with the 
introduction of a elementary scalar field in order to break the electroweak gauge symmetry, leading to an unpleasant system of composite and elementary fields responsible for the gauge symmetry breaking.

\par In Ref.\cite{331-din2} the full  realization of the  dynamical symmetry breaking of a $SU(3)_{{}_{L}}\otimes U(1)_{{}_{X}}$ extension of the SM \cite{tonasse} was  explored. This was accomplished assuming the gauge symmetry $SU(2)_{TC}\otimes SU(3)_{{}_{L}}\otimes U(1)_{{}_{X}}$, where the electroweak symmetry is broken dynamically by a technifermion condensate generated by  the $SU(2)_{TC}$ Technicolor (TC) gauge group, i.e. besides
the exotic $T$ quark condensate and respective composite scalar, we do now have another composite scalar boson formed by $SU(2)_{TC}$ technifermions. 
This symmetry breaking also occurs when we exchange the $SU(2)_{TC}$ group by the $SU(N)_{TC}$ group, as well as when we deal with different
technifermion representations \cite{da}. 

\par In 3-3-1 models where the gauge symmetry breaking is promoted by elementary scalar fields, the many parameters in the scalar potential can
be variated in a large range leaving space to scape, up to now, to the LHC experimental constraints. However, in the case where the gauge symmetry
breaking it totally dynamical, once we describe the possible dynamics of the theory, we may already have some limitation on the possible models.
The study of possible constraints in this class of models is the main motivation of this work. We compute the effective potential for composite operators of a class of 3-3-1 models where the gauge symmetry is dynamically broken, with the main purpose of determining the composite scalar masses. If, for instance, we consider 
$SU(N)_{TC}\equiv SU(2)_{TC}$, we verify that it is quite difficult to generate a scalar boson mass of $125$ GeV, assuming that it is a 
composite scalar that has been observed at the LHC, obeying, at the same time, the lower limit on the $Z^\prime$. 

\par The composite scalar system of these 3-3-1 models have a mixing related to the $Z$ and $Z^\prime$ mixing, which is present in the exotic
$T$ quark self-energy. This mixing will appear in the calculation of the effective potential for composite operators \cite{cjt}, which, when minimized, supply the physical scalar masses, and it is important to know its amount because the scalar masses may be modified by this effect. There are other possible contributions to this mixing, that would appear in an extended theory necessary to explain the
fermion masses, whose spectrum has no explanation in any dynamical symmetry breaking model up to now.  However these corrections are expected to be small
compared to the ones that we discuss here, and it will become clear that already at this level some versions of these models may be excluded by the recent LHC data.

\par The distribution of our paper is the following: In Section II we  present the main aspects of the $SU(N_{TC})\otimes SU(3)_{{}_{L}}\otimes U(1)_{{}_{X}}$  models. In Section III we discuss the mixing of the neutral gauge boson system, its relation to the $T$ quark dynamical mass and present the self-energies that
will be used in the effective action calculation. The mixing in the $T$ quark self-energy  will be responsible by the coupling between the composite scalars ($\phi_{{}_{ T}}$, $\phi$),  associated respectively with the breaking of symmetries $SU(3)_{{}_{L}}\otimes U(1)_{{}_{X}}$  and $SU(2)_{{}_{L}}\otimes U(1)_{{}_{Y}}$  in the  effective potential. Section IV contains the  calculation of the effective action  and in Section V we  compute numerically the scalar boson masses for different TC groups. Finally, in Section VI we draw our conclusions.

\section{$SU(N)_{{}_{TC}}\otimes SU(3)_{{}_{L}}\otimes U(1)_{{}_{X}}$ models}

\par On the lines below we describe the main features of the models, which are similar to that proposed in Ref.\cite{331-din2}, the fermionic content  has the following form
\br 
&& Q_{3L} = \left(\begin{array}{c} t \\ b \\ T  \end{array}\right)_{L}\,\,\sim\,\,({\bf 1}, {\bf 3}, 2/3) \nonumber \\ \nonumber \\
&&t_{R}\,\sim\,({\bf 1}, {\bf 1}, 2/3)\,,\,b_{R}\,\sim\,({\bf 1},{\bf 1},-1/3)\nonumber  \\  \nonumber \\
&&T_{R}\,\sim\, ({\bf 1},{\bf 1},  5/3) \nonumber 
\er
\br
&&Q_{\alpha L} = \left(\begin{array}{c} D \\ u \\ d  \end{array}\right)_{\alpha L}\,\,\sim\,\,({\bf 1}, {\bf 3^*}, -1/3) 
\nonumber \\  \nonumber \\
&&u_{\alpha R}\, \sim\, ({\bf 1},{\bf 1},  2/3 )\,,\,d_{\alpha R}\,\sim\,({\bf 1}, {\bf 1}, -1/3)\nonumber  \\   \nonumber \\
&&D_{\alpha R}\,\sim\, ({\bf 1}, {\bf 1},  -4/3 )
\er
\noindent  where $\alpha = 1,2 $ is the family index and  we represent  the third quark  family  by $Q_{3L}$. In these expressions $({\bf 1},{\bf 3} ,  X)$, $({\bf 1}, {\bf 3^*},  X)$ or $({\bf 1}, {\bf 1},  X)$ denote the  transformation properties  under  $SU(N_{TC})\otimes SU(3)_{{}_{L}}\otimes U(1)_{{}_{X}}$ and $X$ is the  corresponding $U(1)_{X}$ charge.  The leptonic sector includes besides the  conventional  charged leptons and their respective neutrinos the charged heavy leptons $E_a$\cite{tonasse}.  
\br
&& l_{aL} = \left(\begin{array}{c} \nu_{a} \\ l_a \\ E^c_a\end{array}\right)_{L}\,\sim\,({\bf 1},{\bf 3},  0)
\er
\noindent where $a=1,2,3$ is the family index and $l_{aL}$ transforms as triplets   under $SU(3)_L$. Moreover,  we have to add the corresponding right-handed components, $l_{aR} \sim ({\bf 1},{\bf 1}, -1)$ and $E^c_{aR} \sim ({\bf 1},{\bf 1},+1)$.  
\par The fermionic content associated with the TC sector has the form
\br 
&&\Psi_{1L} = \left(\begin{array}{c} U_1\\ D_1\\  U'\end{array}\right)_{L}\,\,\sim\,\,({\bf N_{TC}}, {\bf 3}, 1/2) \nonumber  \\ 
\nonumber \\ 
&&U_{1 R}\, \sim\, ({\bf  N_{TC}}, {\bf 1}, 1/2)\,,\,D_{1 R}\, \sim \,({\bf  N_{TC}}, {\bf 1},-1/2)\nonumber \\
&&U'_{R}\,\sim\, ({\bf  N_{TC}},{\bf 1}, 3/2)\nonumber  ,
\er
\br 
&&\Psi_{2L} = \left(\begin{array}{c} D'\\ U_2\\  D_2 \end{array}\right)_{L}\,\,\sim\,\,({\bf  N_{TC}}, {\bf 3^*}, -1/2) \nonumber \\ 
\nonumber \\ 
&&U_{2 R}\, \sim\, ({\bf  N_{TC}}, {\bf 1}, 1/2)\,,\,D_{2 R}\,\sim\,({\bf  N_{TC}}, {\bf 1},-1/2)\nonumber \\ 
&&D'_{R}\,\sim\, ({\bf  N_{TC}},{\bf 1}, -3/2). \label{3}
\er
\noindent where $1$ and $2$ label the  first and second techniquark families, $U'$ and $D'$ can be considered as exotic techniquarks making an  analogy with quarks $T$ and $D$ that appear in the  ordinary  fermionic content  of the model. The model is anomaly free if we have equal numbers of triplets and antitriplets, counting the color of $SU(3)_c$. Therefore, in order to make the model anomaly free two of the three quark generations transform as ${\bf 3^*}$,  the third quark  family and the three leptons generations  transform as ${\bf 3}$.  It is easy to check that all gauge anomalies cancel out in this model,  in the TC sector the triangular anomaly cancels between the two generations of technifermions.  In the present version of the model the technifermions are singlets of $SU(3)_c$.

\par  As pointed out in Refs.\cite{Das,331-din1}, one interesting feature  of the versions\cite{felice1, frampton, tonasse}  of 3-3-1 models is the following relationship among the coupling constants $g$ and $g'$ associated to the gauge group $SU(3)_{{}_{L}}\otimes U(1)_{{}_{X}}$ 
\begin{equation}
t^2\equiv\frac{\alpha'}{\alpha} = \frac{\alpha_{{}_{X}}}{\alpha} =  \frac{\sin^2\theta_{{}_{W}}(\mu)}{1 - 4\sin^2\theta_{{}_{W}}(\mu)}
\label{eqU1}
\end{equation}
\noindent where $\alpha = g^2/4\pi$, $\alpha' = g'^2/4\pi$   and  $\theta{{}_{W}}$ is the electroweak mixing angle. According to the discussion presented in\cite{Das, 331-din1}, it is precisely this feature of the model that allows  the gauge symmetry breaking  of this version of the $SU(3)_{{}_{L}}\otimes U(1)_{{}_{X}}$  model to  the SM symmetry, because at the scale of a few TeVs   the $U(1)_X$ coupling constant becomes strong as we approach the peak existent in  Eq.(\ref{eqU1}).  Therefore, in the model described in this section the exotic quark $T$  will form a condensate  breaking  $SU(3)_{{}_{L}}\otimes U(1)_{{}_{X}}$  to the electroweak gauge symmetry,  while the SM gauge symmetry will be broken dynamically by a technifermion condensate.  

\par In order to compute the effective action generated for the composite scalar bosons resulting from the two symmetry breaking stages, in the next section we discuss the $T$ quark self-energy, that is related to the mixing between the  standard model  neutral gauge boson Z  with the Z' boson.


\section{The (Z'- Z) mixing and self-energies}

\par In the models that we consider here there is a mixing between the  standard model  neutral gauge boson Z  with the Z' boson,  the mass eigenstates  are\cite{Daniel}  
\br 
&&Z_1 = Z\cos\theta  - Z'\sin\theta  \\ 
&&Z_2 = Z'\cos\theta  + Z\sin\theta 
\er
\noindent where the mixing angle($\theta$) is given by\cite{Daniel} 
\be 
\tan\theta = \frac{M^2_Z - M^2_{Z_1}}{M^2_{Z_2} - M^2_Z},
\ee 
\noindent in this case ($Z_1$) represents the SM neutral boson and ($Z_2$) corresponds to the additional 3-3-1 heavy neutral boson. Therefore, assuming this mixing  
we can write  in the  Euclidean space the following linearized gap equation ($\Sigma_T(p^2)$) for the $T$ quark: 
\begin{eqnarray}
\Sigma_T(p^2) =  &&\,\, a\cos\theta\!\!\int \! dk^2k^2 \frac{\Sigma_T(k^2)}{[k^2 + \mu^2_{X}]}\frac{1}{[(p- k)^2 + M^2_{Z'}]} + \nonumber \\
&&\,\, a\sin\theta\!\!\int \! dk^2k^2 \frac{\Sigma_T(k^2)}{[k^2 + \mu^2_{X}]}\frac{1}{[(p - k)^2 + M^2_{Z}]} \nonumber \\  
\label{10}
\end{eqnarray}
\noindent where
\[
a = \frac{3g^2_{{}_{X}}X_{{}_{L}}X_{{}_{R}}}{16\pi^2},
\]
$\mu_{X}$  is the energy scale where the $U(1)_{X}$ interaction becomes  sufficiently strong to  break dynamically the $SU(3)_{{}_{L}}\otimes U(1)_{{}_{X}}$  to  $SU(2)_{{}_{L}}\otimes U(1)_{{}_{Y}}$, $g^2_{{}_{X}}$ is the $U(1)_{{}_{X}}$ coupling constant, $X_{L}$ and $X_{R}$ are respectively  $U(1)_{{}_{X}}$ charges attributed to the chiral components of the exotic quark  $T$.

\par Besides the condensate and composite states (scalar and pseudoscalar) associated to Eq.(\ref{10}), we have similar entities due to the
$SU(N)_{{}_{TC}}$ group condensation at the scale $\mu_{{}_{TC}}$, generated by a non-trivial technifermion self-energy ($\Sigma^2_{{}_{TC}}(p^2)$). As discussed in Ref.\cite{331-din2}  the  technifermions multiplets, $\Psi_{1}$ and $\Psi_{2}$ described in Eq.(\ref{3}), lead to the formation of composite scalar bosons ($\phi_1$ and $\phi_2$) that are equivalents to the set of fundamental scalar fields, $\rho $ and $\eta $\cite{felice1,tonasse}, so, in order to obtain a  structure of the scalar potential  similar  the described in\cite{felice1,tonasse},  we will assume that   
\be 
\phi = \frac{\phi_1  +  \phi_2 }{\sqrt{2}},
\label{35}
\ee 
\noindent this normalization results from the fact that $f^2_{\pi} \propto \phi^2$ (as we shall describe in Eq.(\ref{28})), and from this  it is possible to verify  that 
\be 
\phi^2 = \phi^2_1 + \phi^2_2, 
\label{38}
\ee 
\noindent once $f^2_{\pi} =  f^2_{\pi_{1}} + f^2_{\pi_{2}}$,  and $\langle\phi_1\rangle = \langle\phi_2\rangle$,
typical of the two technifermion generations of Eq.(\ref{3}). The $f_{\pi_{i}}$ are the technipions decay constants that can be computed through the linearized 
Pagels and Stokar relation\cite{pagels}
\be 
f^2_{\pi} \approx \frac{N_{{}_{TC}}}{4\pi^2}\int\!\!\frac{dp^2p^2\Sigma^2_{{}_{TC}}(p^2)}{(p^2 + \mu^2_{{}_{TC}})^2}
\label{eq34}
\ee
\noindent whereas the pseudoscalar decay constant associated to T quark  self-energy, $\Sigma'_{{}_{T}}(p^2)$  will be written as
\be 
F^2_{{}_{\Pi}} \approx \frac{1}{4\pi^2}\int\!\!\frac{dp^2p^2\Sigma^2_{{}_{T}}(p^2)}{(p^2 + \mu^2_{{}_{X}})^2}. 
\label{eq341}
\ee

\par To compute the effective potential for composite operators \cite{cjt} we need to know the self-energies of the strongly 
interacting fermions: The $T$ quark and the fermions with $SU(N)_{TC}$ charges. Eq.(\ref{10}) has two possible solutions, and
in the program developed in Refs.\cite{soni,soni2} it was verified that the solution falling slowly with the momentum is the
dominating one if suitable new interactions are assumed to be relevant at the scale of the (UV) cutoff. In their case the
gauge boson mass integrals  receive significant contributions from a very large range of loop momenta and the SM gauge boson
masses $M_{W}$ and $M_{Z}$ turn out to be of similar magnitude when compared to the top quark mass. This is exactly the situation 
that we have in the  approach proposed to promote the gauge symmetry breaking of $SU(3)_{{}_{L}}\times U(1)_{{}_{X}}$  to the electroweak 
symmetry, where at the scale of a few TeVs the $U(1)_{{}_{X}}$ coupling constant  becomes strong as we approach the peak existent in Eq.(\ref{eqU1}).
In the Ref.\cite{Das} after the numerical calculation of $M_{{}_{T}}$, it was found that  the magnitudes of $M_{Z'}$ and $M_{{}_{T}}$ are the same order.
Therefore, considering the above comments we  will  assume that the solution of Eq.(\ref{10}) is giving by 
\br
\Sigma_T(p^2) \approx  \Sigma'_T (p^2)\left(1 - h(\omega)\ln\left(\frac{p^2}{\mu^2_{X}}\right)\tan\theta  \right) , \nonumber \\
\label{18}
\er 
and is the one that will be used for the $T$ quark self-energy to determine the  effective potential ($\Omega_{T}$). To write Eq.(\ref{18})
we defined the following quantities:
\[  
\Sigma'_T( p^2) =  \mu_{X}\left(\frac{p^2}{\mu^2_{X}} \right)^{-\left(\frac{1 - \omega}{2}\right)},
\]
where $\omega = \sqrt{1 - 4A}$, $A = a\cos\theta$ and 
\be
h(\omega) = \left( 1 - \frac{1}{\omega} + \frac{A}{\omega}\right) \,\, .
\label{hi}
\ee

In the case of fermions with $SU(N)_{TC}$ charges the self-energy will be given by
\be 
\Sigma (p^2) \sim \mu_{TC} \left[1 + b g^2 \ln\left(p^2/ \mu^2_{TC} \right) \right]^{-\gamma }  \,\,\, ,
\label{eq12}
\ee	
where $\mu_{TC}$ is the $SU(N)_{TC}$ characteristic scale of mass generation,
\be
\gamma= 3c/16\pi^2 b
\label{eq12a} 
\ee
and $c = \frac{1}{2}\left[C_{2}(R_{1}) +  C_{2}(R_{2}) - C_{2}(R_{3})\right] $
where $C_{2}(R_{i})$ are the Casimir operators for fermions in the representations  $R_{1}$ and $R_{2}$ that condense in the representation $R_{3}$,
 $b=(11N-2N_f)/48\pi^2$ for the $SU(N)_{TC}$ group with $N_f$ flavors, $g^2$ is the  coupling constant for which we assume the expression 
\be
{{g}}^2(k^2)= \frac{1}{b \ln[(k^2+4m_g^2)/\Lambda^2]} \, ,
\label{eq4}
\ee
where , $m_g$ is an infrared dynamical gauge boson mass,
whose phenomenologically preferred value is $m_g \approx 2\Lambda $ \cite{cornwall,natale}, and we will set
$\Lambda = \mu_{TC}$. Note that, using the above coupling constant, we are assuming that non-Abelian gauge theories generate
dynamical masses for their gauge bosons \cite{cornwall,aguilar}. As a consequence it is expected that confinement should
be necessary to generate non-trivial fermionic self-energies \cite{cornwall2,dfn}, and the expression for the self-energy
is the one of Eq.(\ref{eq12}), as discussed at length in Ref.\cite{dfn2}. The main features of Eq.(\ref{eq12}), is that it causes the decoupling
of heavier degrees of freedom in models where there is an interaction connecting different fermionic families \cite{dfn2}, it leads to
the deepest minimum of energy, with a vacuum expectation value proportional to $1/g^2$ \cite{natale2,mont,dn1}, it is the only self-energy able to
naturally explain fermion masses as heavy as the top quark \cite{dn3}, as well as the unique possible form of solution that may generate
a light composite scalar boson \cite{dnp2,dln}.

In the self-energies that we discussed above the characteristic scales $\mu_X$ and $\mu_{TC}$ have not been determined up to now. However
they should be constrained by the value of the $SU(3)_{{}_{L}}\otimes U(1)_{{}_{X}}$ gauge boson masses. In order to do so we notice that
for $ M_ {Z '}>> M_Z $, it is possible to show that\cite{Daniel}
\be 
\tan\theta \approx  \frac{1}{2\sqrt{3}t^2}\frac{M^2_{Z}}{M^2_{Z'}} \, ,
\label{19}
\ee 
assuming the result described in Ref.\cite{331-din2} we obtain the masses   
\br
M^2_{Z} &=& \frac{g^2}{4}\left(f^2_{\pi_{1}} + f^2_{\pi_{2}} \right) \left[\frac{1 + 4t^2}{1 + 3t^2}\right] \nonumber \\
&=& \frac{g^2}{4}f^2_{\pi} \left[\frac{1 + 4t^2}{1 + 3t^2}\right] \, ,
\label{191}
\er
\be
M^2_{Z'} =  \frac{g^2}{4}F^2_{\Pi}\left[\frac{4}{3} + 4t^2\right] \, .
\label{20}. 
\ee
\par With these masses and Eq.(\ref{19}), we can write Eq.(\ref{18}) in the form 
\be 
\Sigma_T (p^2) \approx  \Sigma'_T (p^2)\left(1 + A(\omega)\frac{f^2_{\pi}}{F^2_{\Pi}}\ln\left(\frac{p^2}{\mu^2_{X}}\right) \right),
\label{21}
\ee 
\noindent where for $SU(N)_{TC}$
\[
A(\omega) = -\frac{\sqrt{3}{{N}_{TC}}}{16t^2}\frac{\left( 1 + 4t^2\right)}{\left( 1 + 3t^2\right)^2}h(\omega) \, .
\]
The $ f_{\pi}$ decay constant  is related to the SM vacuum expectation value(vev) through
\be
f^2_{\pi} =  \left( f^2_{\pi_{1}} + f^2_{\pi_{2}} \right)  = v^2  = \frac{4M_W^2}{g^2} = (246 GeV)^2.
\ee
\noindent and in the case of $T$ quark self-energy, since there are no evidences for the $Z^\prime$ boson, we just assume $F_{\Pi}\!\sim \!O(\mu_{X})\!\sim\!O(TeV)$. Equations (\ref{eq12}) and (\ref{21}) are the main ingredients to compute the effective action for the model described in section II.

\section{The effective action for composite scalar bosons of the $SU(N_{TC})\otimes SU(3)_{{}_{L}}\otimes U(1)_{{}_{X}}$ model}

\par The effective potential for composite operators~\cite{cjt} is a function of the Green's
functions of the theory, in particular it can be written as a function of the complete fermion ($S$) and gauge boson ($D$) propagators as
\br
V(S,D) &=& - i \, \int \, \frac{d^4p}{(2\pi)^4} \, Tr \left( \ln S_0^{-1} S - S_0^{-1} S +1 \right) \nonumber \\
&&+ V_2 (S,D) \, ,
\label{eq5}
\er
where $S_0$ (and $D_0$) stands for the bare fermion (gauge boson) propagator and $V_2 (S,D)$ is the sum of 
all two-particle irreducible vacuum diagrams. 
The physically meaningful quantity that we must compute is the vacuum energy density
given by
\be
\Omega_V = V(S,D) - V(S_0, D_0) \, ,
\label{eq8}
\ee
where we are subtracting the symmetric part of the potential from the potential
that admits condensation in the scalar channel,  that is denoted by $V(S_0, D_0)$ and is a function of the
perturbative propagators ($S_0$ and $D_0$). 
\par The vacuum energy density, if we remove all indices and integrations, can be 
written as~\cite{cjt,natale2}
\br
\Omega_V &=& - i Tr (\ln S_0^{-1}S - S_0^{-1}S +1) + i Tr \Sigma (S-S_0) \nonumber \\
&+& \frac{1}{2} i Tr (\Gamma S \Gamma S - \Gamma S_0 \Gamma S_0)D  \, .
\label{eq10}
\er
The self-energies described by Eqs.(\ref{eq12}) and (\ref{21}) enter into the definition of $S$.

How the effective action for composite scalar bosons emerge from Eq.(\ref{eq10}) in the case of a dynamically broken gauge
theory including the kinetic term has been detailed by Cornwall and Shellard \cite{cs}, and is also discussed in Refs.\cite{natale2,dnp}.
Here we will skip lengthy details and follow closely the work of Ref.\cite{dnp}, where it was shown that the effective action generated for a TC model could be written in the following way 
\be
\Omega_{TC} = \int d^4x\left[\frac{1}{2Z_{TC}} \partial_{\mu}\phi\partial^{\mu}\phi   - \frac{\lambda_{4_{TC}}}{4}\phi^4 - \frac{\lambda_{6_{TC}}}{6}\phi^6 - ...\right]\,.
\label{23}
\ee

The effective scalar field $\phi(x)$  acts like a dynamical effective scalar field with anomalous
dimension $2\gamma$, is related to the bilinear self-energy $\Sigma (k,p) \propto \phi (k) \Sigma (p)$, it is seem as a variational
parameter, and the kinetic term for our effective theory is obtained inserting $\phi (k)$ in the effective action
and expanding around $k=0$ \cite{cs,natale2,dnp}. In the above equation
\be
Z_{{}_{TC}}  \approx \frac{4 \pi^2 \beta (2\gamma -1)}{N_{TC}N_f} \, ,
\label{eq21}
\ee
\noindent  and 
\br 
&&\lambda_{4_{TC}}= \frac{N_{TC}N_{f}}{4\pi^2}\left(\frac{1}{\beta(4\delta - 1)} + \frac{1}{2}\right) \nonumber \\ 
&&\lambda_{6_{TC}} = -\frac{N_{TC}N_f}{4\pi^2}\left[\frac{1}{\mu^2_{{}_{TC}}}\right].
\label{eqA7}
\er
\noindent In this expressions, $\beta = bg^2$, $N_f$ denote the number of technifermions, 
$\gamma$ has been defined in Eq.(\ref{eq12a}) and is calculated for the respective TC representations.

\par In Eq.(\ref{23}) a term proportional to $\phi^2$ does not appear because we assume that the self-energies are exact solutions of the
linearized gap equations \cite{cjt}, also odd terms in $\phi$ do not appear because we do not have current fermion masses. The constant $Z_ {{}_{TC}}$ arises when the contribution of the kinetic term is included in the calculation of the effective action\cite{cs,natale2, dnp}, and this acts as a normalization constant. This contribution is important in our calculation because it will give the correct normalization of the effective fields, $\phi $ and $\phi_{{}_{T}}$, as discussed in the Refs.\cite{cs,natale2,dnp}. In terms of these fields we can also write the decay constants for the
TC and $T$ fermions as 
\be 
f^2_{\pi} = \frac{N_{{}_{TC}}}{4\pi^2}\frac{\phi^2}{\beta}\frac{1}{(2\gamma - 1)}
\label{28}. 
\ee 
\noindent While for the self-energy, $\Sigma'_{{}_{T}}(p^2)$, we obtain the following relation 
\be 
F^2_{{\Pi}} =  \frac{\phi^2_{T}}{4\pi^2}\frac{1}{2a}
\label{29},
\ee 
\noindent and in this case $\langle \phi_{{}_{T}} \rangle \simeq \mu_{{}_{X}}$. 
\par We can now present the contribution to the effective action due to the composite scalar boson formed by the strong interaction of the exotic $T$ quark. Below we show  the $\phi^4_T$ and $\phi^6_T$ terms of the effective potential $\Omega_T$  and the corrections $\Delta \Omega_T$ assuming the mixture in Eq.(\ref{18}) and the comments leading to Eq.(\ref{21}):
\br 
&&\Omega_{T} = \!\!\int\!\!d^4x\left[\frac{1}{2Z_{T}} \partial_{\mu}\phi_{{}_{T}}\partial^{\mu}\phi_{{}_{T}}   - \frac{\lambda_{4_{T}}}{4}\phi_{{}_{T}}^4 - \frac{\lambda_{6_{T}}}{6}\phi_{{}_{T}}^6 +  \right. \nonumber \\ 
&& \left. \,\,\,\,\,\,\,\,\,\, - \frac{\Delta \lambda_{4_{T}}}{4}\phi_{{}_{T}}^2\phi^2  -  \frac{\Delta \lambda_{6_{T}}}{6}\phi_{{}_{T}}^4\phi^2 +...\right]
\label{eq32}
\er
\noindent where  we identify 
\br
&& Z_{{}_{T}} \approx  8\pi^2 a\,\,\,, \,\,\, \lambda_{4T} \approx \frac{1}{4\pi^2} \left(\frac{1}{4a} + \frac{1}{4} \right) \nonumber \\ 
&& \lambda_{6T} \approx  -\frac{1}{4\pi^2} \frac{1}{\mu^2_{X}} \,\,\,,\,\,\, \Delta \lambda_{4T} = \frac{a\Delta\lambda_4}{\pi^2} \left(\frac{1}{4a} + \frac{1}{4} \right) \\ 
&& \Delta \lambda_{6T} =  -\frac{\Delta\lambda_4}{4\pi^2} \frac{24a^2}{1 + 2a}\frac{1}{\mu^2_{X}} \nonumber \\
&& \Delta\lambda_4 = \frac{\sqrt{3}N_{{}_{TC}}}{16\beta t^2}\left(\frac{1}{(2\gamma -1 )(1-2a)}\right) \nonumber \\
&& \hspace{0.8cm} \times \frac{(1 + 4t^2)}{(1 + 3t^2)^2},
\er
\par In these expressions  we assumed the existence of just one exotic quark that condenses in  the most attractive channel(MAC) \cite{Raby}. 
\par In order to reproduce a standard scalar effective field theory we introduce in our effective Lagrangian the normalized fields 
\be 
\Phi(x) = Z_{{}_{TC}}^{-1/2} \phi(x) \,\, ,
\ee
\be
\Phi_{{}_{T}}(x) = Z_{{}_{T}}^{-1/2} \phi_{{}_{T}}(x) \, .
\ee 
\noindent Now, considering the Eqs.(\ref{35}), where we see that the field $\phi$ actually represent two fields ($\phi_1$ and $\phi_2$), and
adding Eq.(\ref{23}) to Eq.(\ref{eq32}) we can write down the full effective action in terms of the normalized fields $\Phi_{{}_{T}}$
and $\Phi(x)$ (composed by $\Phi_1$ and $\Phi_2$):
\begin{widetext}
\br 
\Omega(\Phi_T, \Phi) = &&\int d^4x\left[\frac{1}{2}\partial_{\mu}\Phi_1\partial^{\mu}\Phi_1 + \frac{1}{2}\partial_{\mu}\Phi_2\partial^{\mu}\Phi_2 + \frac{1}{2} \partial_{\mu}\Phi_{{}_{T}}\partial^{\mu}\Phi_{{}_{T}}  - \frac{\lambda^{R}_{4_{T}}}{4}\Phi_{T}^4 - \frac{\lambda^{R(a)}_{4_{TC}}}{4}\Phi_{1}^4 - \frac{\lambda^{R(a)}_{4_{TC}}}{4}\Phi_{2}^4\right.\nonumber \\ 
&& \hspace*{-0.3cm}\left. - \frac{\lambda^{R(a)}_{4_{TC}}}{4}\Phi_{1}^2\Phi_{2}^2 - \frac{\lambda^{R(a)}_{4_{TC}}}{4}\Phi_{2}^2\Phi_{1}^2 - \frac{\lambda^{R(b)}_{4_{TC}}}{4}\Phi_{T}^2\Phi_{1}^2 - \frac{\lambda^{R(b)}_{4_{TC}}}{4}\Phi_{T}^2\Phi_{2}^2 - \frac{\lambda^{R}_{6_{T}}}{6}\Phi^6_{T}  - \frac{\lambda^{R(a)}_{6_{TC}}}{4}\Phi_{1}^6  - \frac{\lambda^{R(a)}_{6_{TC}}}{4}\Phi_{2}^6 \right. \nonumber \\ 
&& \hspace*{-0.3cm}\left.   - \frac{\lambda^{R(a)}_{6_{TC}}}{4}\Phi_{1}^2\Phi_{2}^4 - \frac{\lambda^{R(a)}_{6_{TC}}}{4}\Phi_{2}^2\Phi_{1}^4 - \frac{\lambda^{R(a)}_{6_{TC}}}{4}\Phi_{1}^2\Phi_{2}^2\Phi_{1}^2  - \frac{\lambda^{R(a)}_{6_{TC}}}{4}\Phi_{2}^2\Phi_{1}^2\Phi_{2}^2 - \frac{\lambda^{R(b)}_{6_{TC}}}{4}\Phi^4_{T}\Phi_{1}^2 - \frac{\lambda^{R(b)}_{6_{TC}}}{4}\Phi^4_{T}\Phi_{2}^2  \right.\nonumber \\
&&\hspace*{-0.3cm} \left.  - \frac{\lambda^{R(b)}_{6_{TC}}}{4}\Phi^4_{T}\Phi_{1}\Phi_{2}  - \frac{\lambda^{R(b)}_{6_{TC}}}{4}\Phi^4_{T}\Phi_{2}\Phi_{1}      \right] 
\label{efp}
\er
\end{widetext}
\par
\noindent where the normalized couplings for the composite fields, $\phi_{{}_{T}}$, $\Phi_1$ and $\Phi_2$ , are given respectively  by 

\br 
&& \lambda^{R}_{4_{T}} =  \frac{Z^2_{T}}{4\pi^2} \left(\frac{1}{4a} + \frac{1}{4}\right) \,\,,\,\,\lambda^{R(a)}_{4_{TC}} = \frac{N_{TC}N_{f}Z^2_{TC}}{4\pi^2} \left(\frac{1}{\beta(4\gamma - 1 )} + \frac{1}{2}\right) \nonumber \\
&& \lambda^{R(b)}_{4_{TC}} = \frac{a\Delta\lambda_4Z_{T}Z_{TC}}{\pi^2} \left(\frac{1}{4a} + \frac{1}{4} \right) \,\,,\,\, \lambda^{R}_{6_{T}} =  -\frac{Z^3_{T}}{4\pi^2} \frac{1}{\mu^2_{X}} \nonumber \\
&& \lambda^{R(a)}_{6_{TC}} = -\frac{N_{TC}N_{f}Z^3_{TC}}{4\pi^2} \frac{1}{\mu^2_{TC}} \,\,\,,\,\,\,  \lambda^{R(b)}_{6_{TC}} = -\frac{\Delta\lambda_4Z^2_{T}Z_{TC}}{\pi^2} \frac{6a^2}{1 + 2a}\frac{1}{\mu^2_{X}}. 
\label{eq-coupl}
\er

\section{The 125 GeV scalar boson  and the  effective action  for the  $SU(N_{TC})\otimes SU(3)_{{}_{L}}\otimes U(1)_{{}_{X}}$ model}

\par To compute the scalar masses  from the effective Lagrangian described in Eq.(\ref{efp}) we use 
\be 
M^{2}_{{}_{\Phi_i}} = \frac{\partial^2\Omega(\Phi_T, \Phi)}{\partial\Phi^2_i}|_{{}_{{}_{\Phi_i=\Phi_{i_{min}}}}} .
\ee 
\noindent where $i= 1,2 $,  and $(1,2)$ label the  first and second techniquark families. After neglecting terms  of higher order 
when substituting the minimum value in the potential we obtain 
\be 
M^{2}_{{}_{\Phi_i}} \approx 2\lambda^{R(a)}_{4_{TC}}\left(\frac{\lambda^{R(a)}_{4_{TC}}}{\lambda^{R(a)}_{6_{TC}}}\right) + 2\lambda^{R(b)}_{4_{TC}}\left(\frac{\lambda^{R(b)}_{4_{TC}}}{\lambda^{R(b)}_{6_{TC}}}\right).
\label{eq41}
\ee 
\par Assuming the above equation, with couplings  $\lambda^{R(a)}_{4_{TC}}$ ,  $\lambda^{R(a)}_{6_{TC}}$ ,  $\lambda^{R(b)}_{4_{TC}}$ and $\lambda^{R(b)}_{6_{TC}}$ defined by Eq.(\ref{eq-coupl}), we finally  can make a scan in the parameter space  $[\mu_{{}_{TC}}\times \mu_{{}_{X}}]$ limiting the Higgs  mass to the range   $ 120GeV < M_H < 130GeV$,    in order to exclude  
possible candidates for TC models.  Note that for this purpose we assume that the ATLAS and CMS are indeed observing a new scalar boson with a mass $M_{\phi} \sim 125$GeV, and consider also the strong limit on the $Z'$  mass  announced by these collaborations\cite{limit-Z'}. 

\par
\begin{figure}[ht]
\centering
\includegraphics[width=0.6\columnwidth]{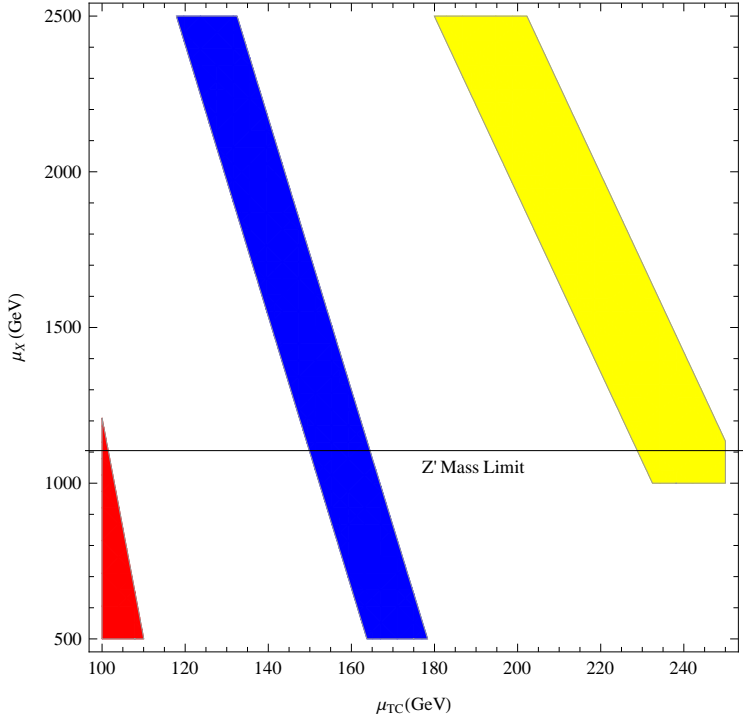}
\vspace*{-0.4cm}
\caption[dummy0]{The region of parameters, $\mu_{{}_{TC}}$ and  $\mu_{{}_{X}}$, for  a composite scalar  mass range  $ 120GeV < M_H < 130GeV$  in the case of $SU(2)_{{}_{TC}}$,  $SU(3)_{{}_{TC}}$  and  $SU(4)_{{}_{TC}}$  TC  groups. The region in red corresponds to  $SU(2)_{{}_{TC}}$, the blue region to $SU(3)_{{}_{TC}}$ and the  yellow region to $SU(4)_{{}_{TC}}$. In this figure we assumed $N_f = 6 $ for all groups. The solid line corresponds to the lower limit on  $\mu_{{}_{X}} $ (i.e. $M_{{}_{Z'}}$).} 
\label{fig-a}
\end{figure}

 The extra $Z'$ boson is predicted in many extensions of the Standard Model at the TeV mass scale,  as in the  Sequential Standard Model $(Z_{ssm})$\cite{Z'2}, with SM like couplings. With LHC data at $(\sqrt{s} = 8 TeV)$,  recently the ATLAS and CMS collaborations placed strong constraints on the mass of these particles\cite{limit-Z'}. 
These constraints depend on the knowledge of the coupling of this boson with SM fermions. In the case of the  $Z_{ssm}$ model with SM like couplings the $Z'$ mass can be excluded below $ 2.49$ TeV. This limit can also be taken as a lower limit on the $Z'$ mass of the models discussed here. In this particular case, if  $M_{{}_{Z'}} >  2.49 TeV$ the energy  scale $\mu_{{}_{X}}$ should be limited to $\mu_{{}_{X}} > 1.1 TeV$. 
In Fig.\ref{fig-a} we  present the allowed region of parameters for  the  $SU(2)_{{}_{TC}}$,  $SU(3)_{{}_{TC}}$  and  $SU(4)_{{}_{TC}}$  cases, with a scalar composite ``Higgs"  mass range  $ 120GeV < M_H < 130GeV$. The solid black line corresponds to the lower limit on $\mu_{{}_{X}}$,  and from this figure we verify that if  $SU(N)_{{}_{TC}}\equiv SU(2)_{{}_{TC}}$ the model barely survives the confrontation with the LHC data.
\par
\begin{figure}[ht]
\centering
\includegraphics[width=0.6\columnwidth]{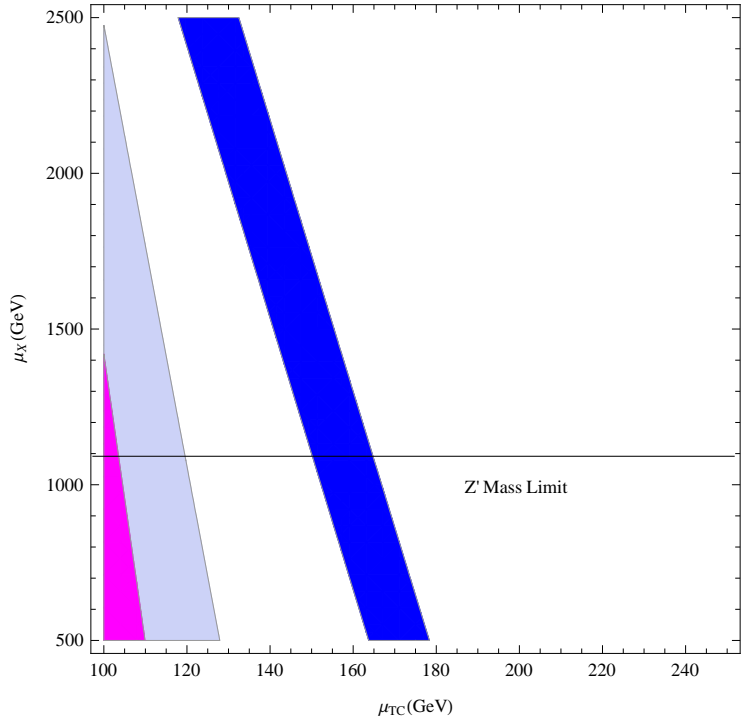}
\caption[dummy0]{The region of parameters, $\mu_{{}_{TC}}$ and  $\mu_{{}_{X}}$, for  a composite scalar  mass range  $ 120GeV < M_H < 130GeV$  in the  $SU(3)_{{}_{TC}}$ case: In this figure the blue region 
corresponds to $N_f = 6 $,  the light blue to $N_f = 8 $ and magenta to $N_f = 10 $.} 
\label{fig-b}
\end{figure}

 In order to complement the analysis,  in the Fig.\ref{fig-b} we show the interval of parameters in the  $SU(3)_{{}_{TC}}$ case assuming 
$N_f = 8 $ and $N_f = 10 $. As can be noticed in the Fig.\ref{fig-b}, the range of values that the parameters $\mu_{{}_{TC}}$ and  $\mu_{{}_{X}}$ can assume decreases when the number of technifermions is increased. The $N_{f}$ dependence on the kinetic term is important and this one is transferred to the $\lambda_6'$s couplings 
and results in a decrease of the scalar mass. 
\par We can write the Eq.(\ref{eq41})  in the following approximate form
\be 
M_{{}_{\Phi_i}}\propto \delta_1 \sqrt{\beta(2\gamma - 1)}\mu_{{}_{TC}} + \delta_2 \sqrt{\beta(2\gamma - 1)}\mu_{{}_{X}}
\ee 
\noindent where $\delta_1$ and $\delta_2$  are constants.  The increase of the number of fermions implies in an increase  of $\sqrt{\beta(2\gamma - 1)}$, then to keep the interval
to the Higgs  mass  $ 120GeV < M_H < 130GeV$   we observe a decrease in the area $[\mu_{{}_{TC}}\times \mu_{{}_{X}}]$    in the parameter space. This 
effect is the same that appears in the normalization of the Bethe-Salpeter wave function discussed in Ref.\cite{dnp2}.   

\par
\begin{figure}[t]
\centering
\includegraphics[width=0.6\columnwidth]{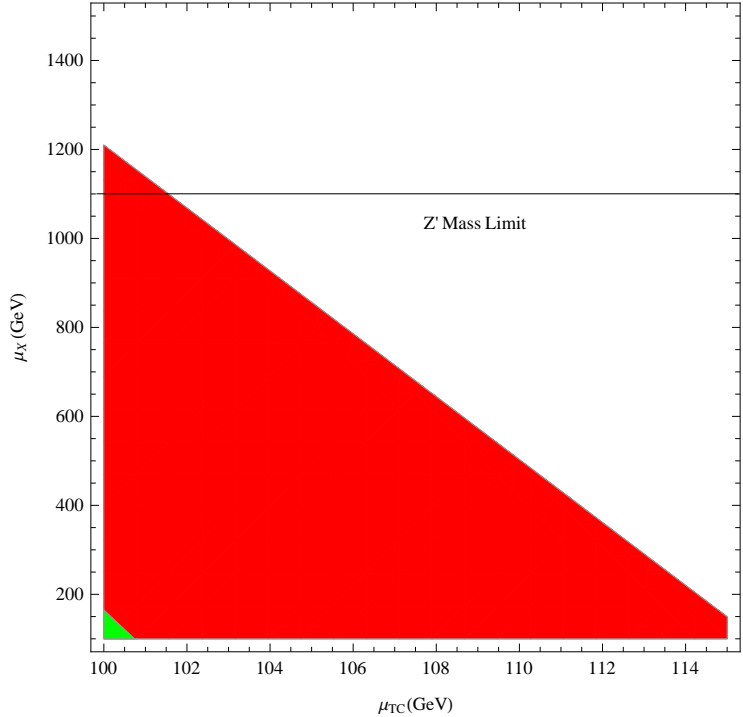}
\caption[dummy0]{The region of parameters, $\mu_{{}_{TC}}$ and  $\mu_{{}_{X}}$, for a composite scalar  mass range  $ 120GeV < M_H < 130GeV$  in the $SU(2)_{{}_{TC}}$ case: The small Green region  corresponds to $N_f = 7 $, and the solid line corresponds to the lower limit on  $\mu_{{}_{X}}$.  } 
\label{fig-c}
\end{figure}

 In the  Fig.\ref{fig-c} we show the interval of parameters in the $SU(2)_{{}_{TC}}$ case assuming  $N_f = 7 $ (in Green), and again we  verify that  $SU(2)_{{}_{TC}}$ case with this number of fermions can be  ruled out in confrontation with the LHC data.

\section{Conclusions} 


\par In this work we computed an effective action for the composite scalar boson system, $\phi_{{}_{T}}$, $\phi_1$ and $\phi_2$, formed by the fermions and technifermions $Q_{{}_{3}}$, $\Psi_1$ and $\Psi_2$  described in the  Eqs.(1) and (3). We include in this calculation the kinetic term of the effective theory. This term is important because it provides a normalization factor for the effective scalar boson Lagrangian. The effective Lagrangian is then normalized in order to reproduce a standard scalar effective field theory, leading to a non-trivial set of scalar self-couplings. From this Lagrangian we can determine the scalar boson masses of the theory.
 
\par To compute the effective action for the model described in the Section II, we firstly determined the correction to the $T$ quark self-energy that result from the mixing between the  standard model  neutral gauge boson $Z$  with the $Z^\prime$ boson. We show that this correction is the responsible by introduce the coupling between the composite scalars,  $\phi_{{}_{ T}}$ and  $\phi_{1,2}$  associated respectively to the $T$ quark and to the technifermions. In Section III we discussed the self-energies used to compute the  effective potential ($\Omega_{T}$), where we assumed that the interaction $U(1)_{{}_{X}}$ plays a role analogous to the ultraviolet dynamical symmetry breaking program proposed in Ref.\cite{soni,soni2}. We also assume a TC self-energy that decays slowly  with the momentum whose origin is due to the introduction of confinement in the gap equation, as discussed in Ref.\cite{dfn2}, that is typical of the gauged Nambu-Jona-Lasinio (NJL) type of models where the anomalous dimension is $\gamma_m \approx 2$. Note that it is hardly possible to generate light scalar
composite bosons without this particular choice \cite{dnp2,dln}. We finally determined an effective scalar boson Lagrangian and from it we obtained the
scalar boson masses associated to these models.

\par As already discussed in the introduction, the models that we consider here are interesting due to the particular form of their anomaly cancellation, due to the fact that it is one example of model where we have a naturally strong Abelian theory at the TeV scale, capable of producing a dynamical symmetry breaking
of the model to the SM symmetry, whereas the SM gauge symmetry is broken by a TC condensate. Within this class of models we can also explain the larger
decay rate of the $125$ GeV boson into photons that is observed by the LHC experimental groups, caused by the presence of  extra charged vector boson($V^+$, $U^{++}$) \cite{PA}. The only ingredients of our calculation are the self-energies of the strongly interacting theories. From these ones we can compute
the scalar masses and the neutral gauge boson masses. Assuming that the $125$ GeV observed at the LHC is a scalar boson and using their lower limit in
the $Z^\prime$ mass, we compare the experimental results with the mass values that we obtained for these quantities. We verified that the models are
strongly constrained, showing that it is rather difficult to have light scalar composites and at the same time to generate masses for neutral gauge bosons where one of them (the $Z^\prime$) is quite heavy. It is possible that in models with the presence of fundamental scalar bosons such difficulty is not present
due to the many parameters that can be adjusted in the scalar potential, however, in this case, we may also foresee that this adjustment may lead to unnatural
values of the coupling constants, unless some discrete symmetries are introduced by hand in order to avoid undesirable terms in the scalar potential.

\section*{Acknowledgments}
This research was  partially supported by the Conselho Nacional de Desenvolvimento Cient\'{\i}fico e Tecnol\'ogico (CNPq).

\end{document}